# Customized spin spirals in ferromagnetic thin films

Anjali Panchwanee,[1, a)] Kai Schlage,[1, b)] Dieter Lott,[2] Sven Velten,[1] Thomas Saerbeck,[3] David L. Cortie,[4] Sakshath Sadashivaiah,[5] Ilya Sergeev,[1] Guido Meier,[6, 7] Lars Bocklage,[1, 7] and Ralf Röhlsberger [1, 5, 7, 8, 9]

1) Deutsches Elektronen-Synchrotron DESY, Notkestraße 85, 22607 Hamburg, Germany

2) Institute for Materials Research, Helmholtz-Zentrum Geesthacht, Max-Planck-Straße 1, 21502 Geesthacht, Germany

3) Institut Laue-Langevin, 71 avenue des Martyrs, CS 20156, 38042 Grenoble cedex 9, France

4) Australian Nuclear Science and Technology Organisation, New Illawarra Road, Lucas Heights, NSW 2234, Australia

5) Helmholtz-Institut Jena, Fröbelstieg 3, 07743 Jena, Germany

6) Max Planck Institute for the Structure and Dynamics of Matter, Luruper Chaussee 149, 22761 Hamburg, Germany

7) The Hamburg Centre for Ultrafast Imaging, Luruper Chaussee 149, 22761 Hamburg, Germany

8) Institut für Optik und Quantenelektronik, Friedrich-Schiller-Universität Jena, Max-Wien-Platz 1, 07743 Jena, Germany

9) GSI Helmholtzzentrum für Schwerionenforschung GmbH, Planckstraße 1, 64291 Darmstadt, Germany

The advancement of spintronic nanoscale devices hinges on the ability to flexibly engineer magnetic spin structures in thin-film stacks with precision and control. Meeting this demand remains a challenge for stable non-collinear spin configurations and, more specifically, vertical spin spirals in thin films. Innovative methods are required for their fabrication, stabilization and control. Here, we use oblique-incidence deposition to design and stabilize vertical spin spirals at room temperature and without an external field in magnetic thin films. We induce two crossed uniaxial magnetic anisotropies at the thin film boundaries. These anisotropies are tunable in direction and strength, thus providing control over the angular range and depth profile of the resulting spin spiral. The combination of polarized neutron reflectometry and nuclear resonant scattering enables precise and direct determination of the depth-dependent spin configurations. Our results establish a single-film design approach, in which the surface anisotropies independently serve as controllable design parameters for tailoring the vertical spin-spiral profile. Potential applications include nanoscale energy-storage devices, magnetic sensors, and ferromagnetic-resonance filters, advancing all-spin-based device engineering in general.

## 1. Introduction

Engineering of synthetic magnetic helices has attracted growing attention in recent years as a route to unlock novel functionalities for next-generation data storage, spintronic devices, and magnetic sensors.[1–7] In particular, vertical spin-spiral structures in thin films, and their ability to adopt multiple spin configurations provide an additional degree of freedom for tailoring magnetic states, which is important for all-spin-based device applications.[9] In this context, the concept of energy-storing elements that utilize spin at every stage of operation has been theoretically proposed in modulated helices with an integer number of twists by Vedmedenko et al.,[10], where the corresponding magnetic states can remain topologically stable even in the absence of chiral Dzyaloshinskii-Moriya interaction. By incorporating boundary constraints such as magnetic anisotropy including magneto-crystalline, shape and surface/interface anisotropy together with intrinsic magnetic interactions like exchange, Ruderman-Kittel-Kasuya-Yosida (RKKY), or long-range dipolar interactions, non-collinear spin structures such as vertical magnetic spin spirals and helices, can be stabilized in thin films and multilayers.[9,11–14]

Various approaches have been explored to fabricate and stabilize magnetic spin spirals. A spin-spiral state was experimentally demonstrated in conventional exchange-spring bilayer systems. [15] However, the spin spiral was not stable at zero field and required an external magnetic field for stabilization. Alternatively, rare-earth (RE)-Fe and RE-RE-based multilayers have been employed to fabricate and stabilize magnetic spin spirals by exploiting the strong in-plane magnetocrystalline anisotropy of rare-earth elements and their exchange coupling with adjacent Fe layers. [16–18]For example, Fust et al.[16] topologically stabilized spin helices in Dy-Fe multilayers at low temperature (15 K), where the strong in-plane magnetocrystalline anisotropy of Dy and the exchange coupling across the Fe-Dy interfaces played key roles. These approaches, however, rely on the specific material combinations and, in some cases, cryogenic temperatures, which limit their practical applicability. It has also been theoretically[8] demonstrated that vertical magnetic helices can be stabilized at zero field in exchange-coupled trilayer structures, where a magnetically soft Fe layer is confined between two hard magnetic layers with different anisotropies. In such systems, the hard magnetic layers impose distinct magnetic constraints at the two boundaries, while the exchange coupling accommodates these constraints through a continuous rotation of the magnetization in the soft layer. This theoretical study highlights that the precise control of the magnetic anisotropy, and hence of the associated magnetic boundary conditions, is critical for the formation and stabilization of the vertical spin spirals. However, experimentally achieving such controlled magnetic boundary conditions remains challenging, and realizations of the resulting spin-spiral states at room temperature and zero field are still scarce.

Oblique incidence deposition (OID) has emerged as a versatile and widely used route for setting and controlling uniaxial magnetic anisotropy in thin films by inducing nanoscale wavy surface morphologies through the self-shadowing effect during deposition.[19–22] By controlling the azimuthal ($\alpha$) and polar ($\theta$) orientations of the atomic flux impinging on the surface [23,23](see Fig.1(a)), both the orientation and strength of the resulting in-plane uniaxial anisotropy can be precisely tailored. This approach has been used to generate non-collinear vertical spin profiles in multilayer structures and to enhance magnetoresistive properties.[24] It has subsequently enabled flexible tuning of thin-film tunneling magnetoresistance (TMR) devices.[24] In addition, OID-induced enhancement of uniaxial magnetic anisotropy in THz emitters has been shown to enable stable THz emission without an external magnetic field.[25]

Here, we present a novel fabrication approach for stabilizing and tailoring vertical spin-spiral profiles at room temperature and zero field in polycrystalline magnetic thin films using sequential oblique incidence deposition (OID). [23,26] The method imprints two crossed uniaxial magnetic anisotropies with independently tunable orientations and strengths at the bottom and top surfaces of a single magnetic thin film through nanoscale modification of the surface morphology. These anisotropies impose distinct magnetic boundary conditions across the film thickness. After saturation in an external magnetic field applied at a defined angle relative to the imprinted easy axes and subsequent field removal, the magnetic moments near the two surfaces relax toward their respective easy-axis directions, while the magnetization within the film rotates continuously through the thickness, thereby stabilizing a vertical spin-spiral at zero field. By systematically varying the strength of one surface anisotropy while maintaining a fixed relative orientation between the two easy axes, we investigate how the depth-dependent spin profile evolves with surface anisotropy strength. The depth-dependent magnetic configurations are characterized using complementary polarized neutron reflectometry (PNR) and nuclear forward scattering (NFS) of synchrotron radiation. Rather than relying on chemically distinct magnetic multilayers, complex interlayer-coupling schemes, cryogenic conditions, or external-field stabilization, the present approach realizes boundary-engineered spin spirals within a single continuous ferromagnetic Fe film. In this way, the orientation and strength of the anisotropies at its two opposing interfaces become design parameters for tailoring the depth profile of the non-collinear magnetic state.

## 2. Results

### 2.1 Engineering crossed surface anisotropies in a continuous film via oblique-incidence deposition

In a thin polycrystalline film deposited at oblique incidence, elongated grains form through a self-shadowing effect during the initial and subsequent stages of growth.[22 27]The amplitude of the resulting shape-induced surface waviness, and consequently the strength of the uniaxial anisotropy, increase with both the polar deposition angle (*θ)* and the film thickness (t). This effect enables a continuous tuning of the coercivity, from the non-OID value up to several tens of milliTesla[28], as illustrated in Figs. 1(c, e) for non-OID and OID Fe films, respectively.

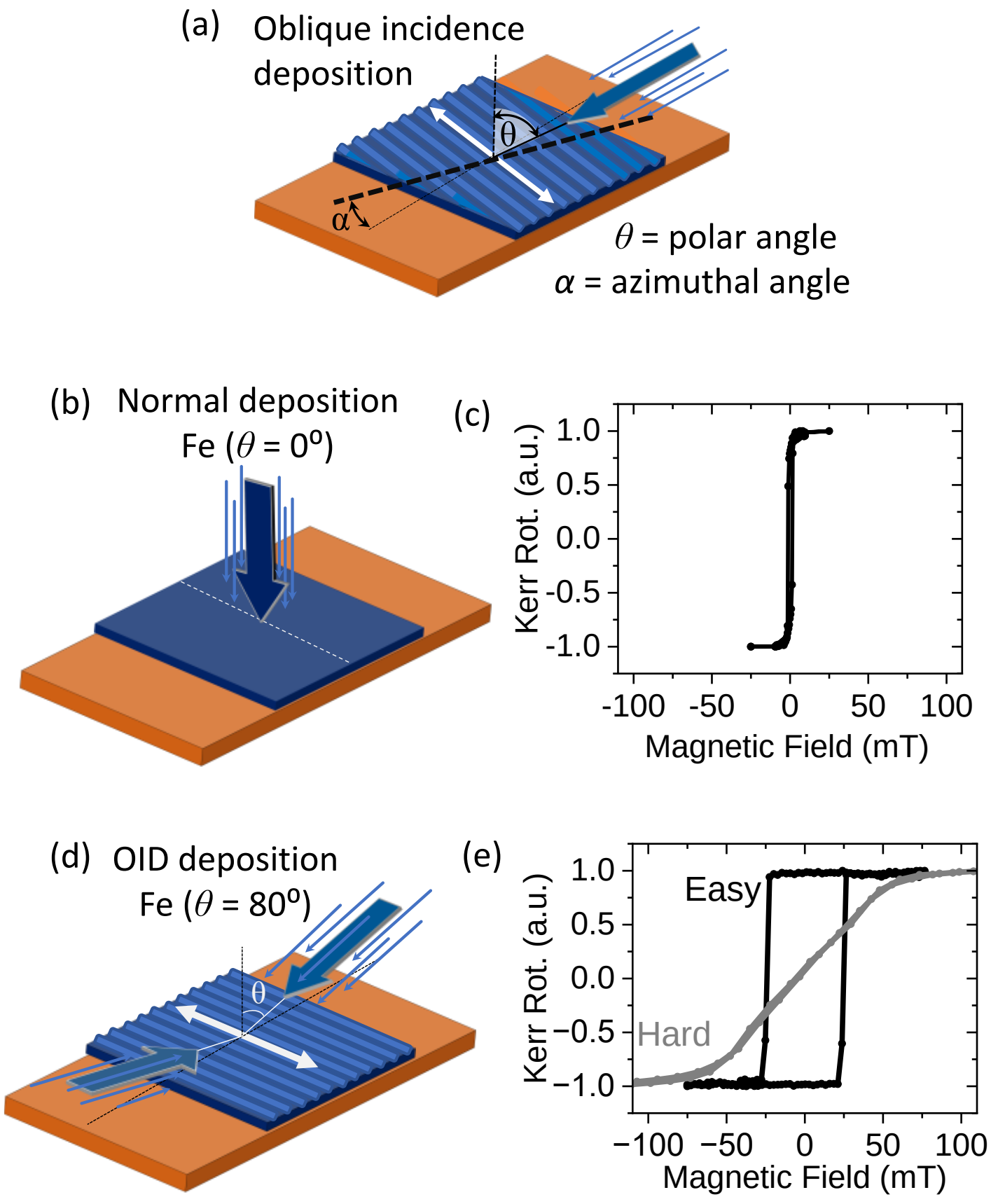

Fig. 1. Control of in-plane uniaxial magnetic anisotropy in Fe thin films by oblique incidence deposition (OID). (a) Schematic of the OID approach. The atomic flux reaches the substrate at a defined azimuthal orientation α and a polar angle *θ*, resulting in a thin film with a wavy surface profile (light-blue lines). Normal deposition and corresponding magnetic hysteresis in panels (b, c) for a 5 nm Fe film deposited at $\theta = 0^\circ$, showing no preferred in-plane magnetic

orientation, and OID deposition and corresponding magnetic hysteresis in panels (d, e) for a Fe film deposited at $\theta = 80^{\circ}$. The OID film shows a pronounced uniaxial anisotropy with the easy axis oriented perpendicular to deposition direction, as indicated by the white arrow, and a hard magnetic axis parallel to the deposition direction.

An Fe film deposited at $\theta = 0$ exhibits no preferred in-plane magnetic orientation in magneto-optic Kerr effect (MOKE) measurements (Fig. 1(c)). In contrast, deposition at 80˚ induces a uniaxial anisotropy in the Fe layer, as evidenced by the distinct easy- and hard-axis MOKE loops in Fig. 1(e), with the easy axis oriented perpendicular to the azimuthal deposition direction. Since OID from a single direction produces a thickness gradient,[29] this gradient is minimized by deposition from two opposite azimuthal directions, as indicated by the blue arrows in Fig. 1 (d). The OID Fe film exhibits elongated grains, as shown by atomic force microscopy (AFM) topography in the supplementary Fig. S1. This morphology is consistent with the uniaxial magnetic anisotropy induced by OID, whose strength and coercivity can be tuned by varying the polar deposition angle

A well-defined easy axis can also be imprinted in a normally ($\theta = 0^{\circ}$) deposited Fe film using a non-magnetic OID underlayer. This is demonstrated for a 10.4 nm Pt layer deposited at $\theta = 80^{\circ}$, followed by a 5 nm Fe film (Fig. 2 (a)). The obliquely deposited Pt layer forms elongated, connected Pt columns oriented perpendicular to the Pt deposition direction, producing a wavy surface morphology (Fig. 2 (b)). This morphology is transferred to the normally deposited Fe layer, thereby imprinting a distinct magnetic easy axis (Fig. 2 (c)). Together, Figs. 1 and 2 demonstrate that OID enables independent control of the orientation and relative strengths of the anisotropies associated with the top and bottom surfaces of a single magnetic film, providing the basis for realizing controlled crossed uniaxial anisotropy axes within the same magnetic layer. We refer to this implementation as the double-sided OID approach.

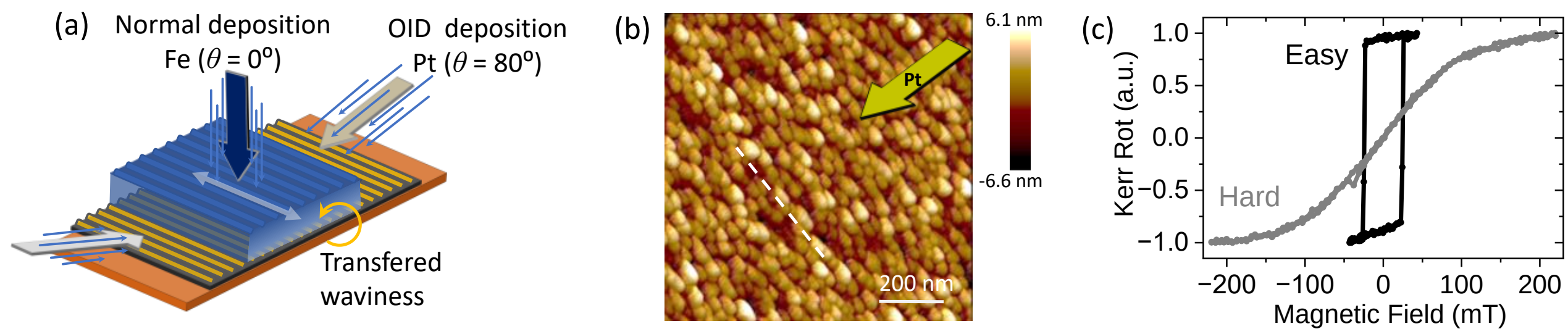

Fig. 2 OID-enabled transfer of magnetic anisotropy through a non-magnetic Pt underlayer. (a) Deposition of a 10.4 nm non-magnetic Pt layer at $\theta = 80^{\circ}$ from two opposite azimuthal directions. (b) AFM topography of the 10.4 nm Pt layer, showing a surface-waviness amplitude of 1.8 nm and a period of 24 nm. The yellow arrow indicates the Pt deposition direction, while the white dotted line marks the coalescence of Pt islands perpendicular to the incident atomic flux, resulting in elongated Pt islands/stripes. (c) Magnetic hysteresis of a 5 nm Fe layer deposited at normal incidence on top of the OID Pt layer. The distinct easy- and hard- axis responses demonstrate the imprinting of a magnetic easy axis in the Fe layer through the transferred surface waviness of the underlying OID Pt film.

Using this approach, we fabricate a 100 nm Fe film with crossed top and bottom easy axes. For the relative easy-axis orientation chosen in this study, the two axes define complementary in-plane angular separations of 80˚ and 100˚. This arrangement allows, in principle, up to four remanent spin-spiral configurations: two associated with an 80˚ and two with a 100˚ opening angle. Depending on the magnetic-field orientation, one of these configurations can be stabilized. In the present study, we focus on the configuration with an intended opening angle of 100˚.

The preparation steps used to realize this sample structure are schematically illustrated in Fig. 3. First, a uniform 8 nm non-magnetic OID Pt layer is deposited on a Si substrate at a polar angle of $\theta = 80^{\circ}$ ($Pt_{80^{\circ}}$) (Fig. 3(a)). Subsequently, a 92 nm Fe layer is deposited at normal incidence ($Fe_{0^{\circ}}$) on top of the OID Pt layer (Fig. 3(b)). The wavy morphology of the Pt layer is transferred to the bottom interface of the Fe layer, thereby imprinting the bottom magnetic easy axis (BEA; white arrow) through correlated roughness. The sample is then rotated azimuthally by α = 100˚ anticlockwise from the BEA direction, followed by oblique deposition of an 8 nm uniform Fe layer at $\theta$ = 80˚ ($Fe_{80°}$) (Fig. 3(c)). This final deposition step establishes the top uniaxial magnetic anisotropy, as indicated by the upper white arrow in Fig. 3(c). Owing to the uniaxial nature of the two easy axes, their relative orientation corresponds to an angular separation of 100˚ when measured anticlockwise from the BEA and, equivalently, 80˚ in the opposite direction, consistent with the complementary spin-spiral opening angles introduced above. Separate reference films yield uniaxial anisotropy energy densities of $K_u$ = 5.25 x $10^4$ J/$m^3$ for the bottom-anisotropy conditions and 3.8 x $10^4$ J/$m^3$ for the OID-Fe top-anisotropy conditions. These values are therefore used as estimates for the corresponding anisotropy strengths in the double-sided OID sample structure used in this study. The intended 100˚ spin-spiral configuration is schematically illustrated in Fig. 3(d).

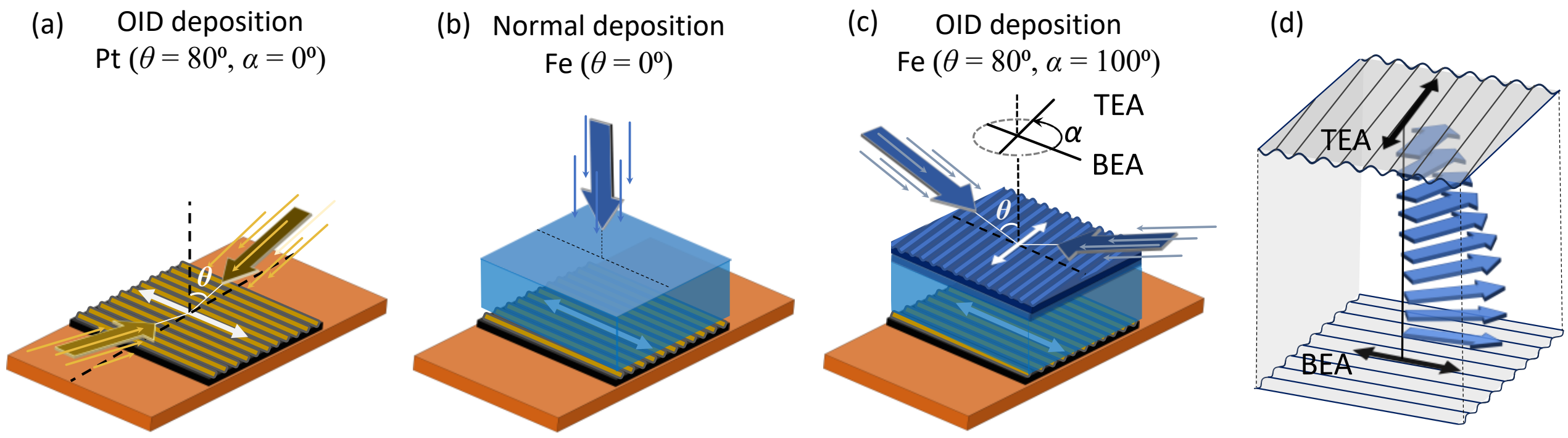

Fig. 3. Double-sided OID fabrication of the spin-spiral sample. (a) Formation of the wavy Pt template by deposition of an 8 nm Pt layer at $\theta$ = 80˚. (b) Deposition of a 92 nm Fe layer at normal incidence on top of the OID Pt layer, transferring the Pt surface morphology to the bottom interface of the Fe layer and imprinting the bottom easy axis (BEA), indicated by the white bidirectional arrow. (c) OID deposition of an 8 nm Fe layer at $\theta$ = 80˚ after an azimuthal rotation of α = 100˚, establishing the top easy axis (TEA) relative to the BEA. (d) Schematic illustration of the intended 100˚ spin-spiral configuration at remanence.

## 2.2 Depth-dependent spin profile determination

Polarized neutron reflectivity (PNR) was employed to probe the magnetic depth profile and spin structure of the sample shown in Fig. 3 (c). By measuring the reflected neutron intensity in all four polarization channels ($R^{++}$, $R^{--}$, $R^{+-}$, and $R^{-+}$) as a function of the momentum transfer $q_z$, PNR provides sensitivity to both the structural and magnetic depth profiles of the sample. Polarization analysis provides information on the in-plane components of the magnetic vector ($m_x$, $m_y$) as a function of film depth, enabling the identification of depth-dependent magnetic spin configurations.

Room-temperature PNR measurements were performed in a high-field state under an external magnetic field of $\mu_0H$ = 110 mT applied along the x-axis, with the sample oriented as shown in Fig. 4(a). The experimental and fitted reflectivity is shown in Fig. 4(b), and the corresponding spin asymmetry, SA= ($R^{++}$ - $R^{--}$) / ($R^{++}$ + $R^{--}$), is presented in Fig. 4(c). The fitted depth-dependent magnetic scattering-length density (SLD) components along the x- and y-directions are represented by $m_x$ and $m_y$, respectively, in Fig. 4(d). The $m_x$ profile shows that the magnetization is predominantly aligned along the applied field throughout the film thickness, while small deviations of approximately 5° near the top and bottom interfaces are inferred from the finite transverse magnetic component $m_y$. This indicates that the sample is not fully magnetically saturated at 110 mT field and the spins at the top and bottom surfaces are not fully aligned with the external field (Fig.4(e)) as expected for this iron film. This is consistent with the SQUID measurements, which likewise show that the magnetic saturation is not reached at this field (see Supplementary information). A magnetic moment of 1.3 $\mu_B$/Fe atom is obtained from the PNR fit and SQUID measurements at 110 mT. The corresponding top-view representation of the fitted spin distribution through the film thickness is shown in Fig. 4(e).

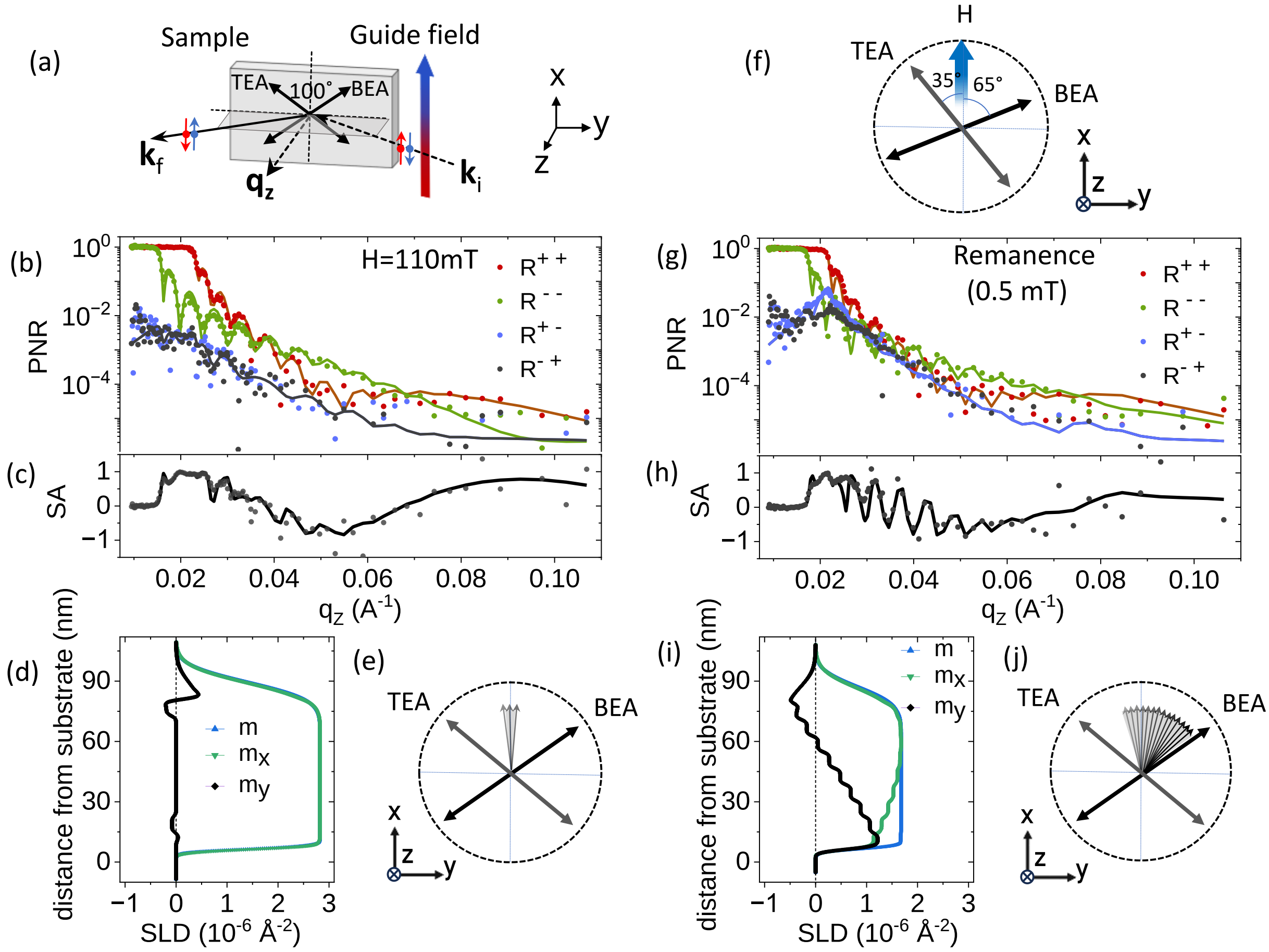

Fig. 4. Polarized neutron reflectometry of the 100 nm Fe spin-spiral sample. Top-view schematics of the PNR scattering geometries for (a) the high-field measurements and (f) remanent state measurements. (b, g) Experimental PNR data (symbols) and fitted reflectivities (solid lines) for the non-spin-flip ($R^{++}$, $R^{--}$), and spin-flip ($R^{+-}$, $R^{-+}$) channels. (b) PNR measured under an applied field of 110 mT and (g) at remanence under a 0.5 mT neutron guiding field. Experimental and fitted SA are shown for (c) the high-field state and (h) the remanent state. Extracted magnetic SLD depth profiles of the in-plane magnetization components $m_x$ and $m_y$, together with the magnetization magnitude m, are shown for (d) the high-field state and (i) the remanent state. In (d), the m and $m_x$ profiles overlap, and therefore only $m_x$ is visibly distinguishable. (e, j) Top-view representations of the fitted depth-dependent spin configurations in the high-field and remanent states, respectively. The remanent state profile shows the formation of a spin spiral.

To induce and probe the spin spiral, we measured the reflectivity at remanence. An external magnetic field of $\mu_0H$ = 110 mT was first applied along the x-direction, corresponding to an angle of 35° relative to the TEA, as shown in Fig. 4(f). The field was subsequently reduced to a small guiding field of 0.5 mT (remanent state). This guiding field was sufficient to maintain the quantization axis of the polarized neutrons along x-axis but does not influence the Fe magnetization orientation in the sample. Upon field reduction, the Fe magnetization is expected to relax from the field-aligned state toward the imprinted top and bottom easy axis directions, potentially giving rise to a depth-dependent spin rotation, as schematically illustrated in Fig. 3(d). The remanent-state reflectivity shows finite spin flip (SF) contributions ($R^{+-}$, $R^{-+}$) in Fig. 4(g), indicating a magnetization component transverse to the neutron polarization axis along the depth. The depth dependence of the magnetization was determined by fitting the four polarization channels.

For the PNR analysis, the depth-dependent magnetic state was described using a uniform-pitch spiral model, in which the magnetization magnitude is kept constant and the magnetization direction changes by an equal angular increment between successive sublayers. The fitted reflectivity and the corresponding SA are shown in Figs. 4(g) and 4(h), respectively, while the resulting $m_x$ and $m_y$ depth profiles are presented in Fig. 4(i). The $m_y$ depth profile reveals that the magnetization in the lower part of the Fe layer is oriented closer to the BEA and gradually rotates through the film thickness toward the crossed TEA, without reaching complete alignment with the TEA. The reconstructed profile gives a spin-spiral configuration with an opening angle of approximately 68˚, as illustrated schematically in Fig. 4(j). The incomplete rotation toward the TEA, together with the different top- and bottom-anisotropy strengths introduced above, raises the question of how the relative surface-anisotropy strengths influences the depth-dependent spin profile.

## 2.3 Spin profile tuning via the bottom magnetic anisotropy

PNR already provides a strong signature of the depth-dependent spin-spiral within the film. However, the exact profile depends on the model assumed and the quality of the fit of the signal reflected from the total thickness. In order to get a view of the spin profile at a specific depth, we use nuclear forward scattering (NFS) of the 14.4 keV nuclear transition in $^{57}$Fe. The NFS signal is sensitive to the magnetization direction relative to the incident x-ray beam and its polarization. By using a thin isotopically enriched $^{57}$Fe probe layer within the Fe film, this technique provides the local magnetic hyperfine field and magnetization vector, enabling a corresponding probe-layer depth view of the spiral [15,19,30]

To examine the role of the anisotropy strength in determining the spiral opening angle and its depth profile, we prepared a dedicated sample having a wedged OID Pt layer with a linear thickness gradient. This layer varies from 5.4 (thin-Pt) nm to 10.4 nm (thick-Pt) as indicated in Fig. 5 (a). It provides a systematic variation of the bottom-anisotropy strength across the sample by increasing the amplitude of the surface waviness and the concomitant induced uniaxial anisotropy from $K_u$ = 4x10$^4$ J/m$^3$ at $t_{Pt}$ = 5.4 nm to 6.6 x10$^4$ J/m$^3$ at $t_{Pt}$ = 10.4 nm (Supplementary Figs. S1 and S2). For the top easy axis prepared by OID, a 90° orientation to the bottom easy axes is used. Its anisotropy is approximately the same over the wedge of the sample. However, correlated roughness transferred from the bottom film can be present at the top layer. The 10 nm isotopically enriched $^{57}$Fe probe layer (yellow) was embedded diagonally within the Fe film of 100 nm total thickness as shown in Figs. 5(b) and (c). This wedged and isotopically enriched sample provides a systematically variation of the bottom-anisotropy strength while keeping the top-anisotropy condition approximately fixed.

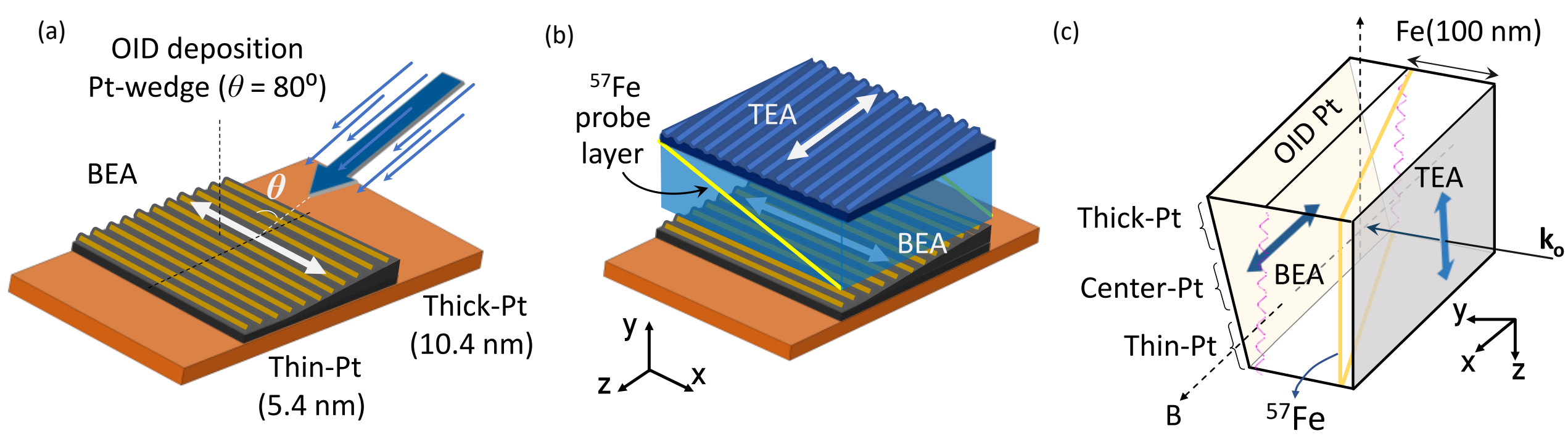

Fig. 5. Wedged Pt layer and $^{57}$Fe probe-layer design for depth-resolved NFS measurements. (a) A OID geometry is used to prepare the Pt wedge at a polar angle of $\theta$ = 80˚, with the thickness varying from 5.4 nm (thin-Pt) to 10.4 nm (thick-Pt). The bottom easy axis (BEA) is indicated by the white arrow. (b) Schematic of the 100 nm Fe sample containing the diagonally embedded $^{57}$Fe probe layer and crossed TEA and BEA. (c) NFS measurement geometry showing the thin-, center-, and thick-Pt regions and the position of the $^{57}$Fe probe layer within the Fe film. In the local coordinate system shown in panel (c), the x-direction denotes lateral translation at fixed Pt-wedge thickness, whereas the z-direction denotes translation across the Pt wedge, corresponding to a change in Pt thickness and bottom-anisotropy condition. The blue arrows indicate the TEA and BEA directions. $k_0$ denotes the incident x-ray wavevector.

Within the continuous Pt wedge, three regions: thin-Pt, center-Pt, and thick-Pt were selected for the NFS measurements. The center-Pt region ($t_{Pt}$=8 nm) was chosen to match the Pt thickness and bottom-anisotropy condition used in the PNR spin-spiral sample. The thin- and thick-Pt regions provide systematically weaker and stronger bottom-anisotropy conditions, respectively. At each Pt-thickness region, the sample is translated laterally along the direction of constant Pt thickness (x-direction) and therefore constant bottom-anisotropy condition (Fig. 5(c)). Because the $^{57}$Fe probe layer is embedded diagonally through the Fe film, this lateral translation positions probes different depths. Please note that we probe about 70 nm of the total iron film because of the finite marker-layer thickness, the experimental geometry, and the preparation of the topmost 20 nm and bottommost 15 nm OID layers with non-enriched iron.

Prior to the NFS measurements, the sample was saturated in an external magnetic field of 230 mT applied at an angle of 5° relative to the TEA. Fig. 6(a) shows a top-view schematic of the sample with the TEA and BEA oriented 90° apart and the applied saturation field. Simulated NFS quantum-beat patterns for fully in-plane magnetization aligned along either the TEA or the BEA are plotted in Fig. 6(b). Intermediate in-plane magnetization orientations produce a gradual evolution between these two cases. In addition, an out-of-plane magnetization component produces an additional characteristic modification of the quantum-beat pattern, allowing out-of-plane moments to be identified by NFS. Measurements were performed at remanence in the thin-, center-, and thick-Pt regions The resulting depth-dependent NFS spectra for the three bottom-anisotropy conditions are shown in Figs. 6(c) to (e). In all three regions, the quantum-beat patterns exhibit characteristic changes with probe-layer depth, indicating a gradual rotation of the magnetization vector ($\vec{m}$) from the BEA toward the TEA. These observations indicate the formation of a depth-dependent spin-spiral structure in all three Pt regions. The evolution of the spectra differs among the thin-, center-, and thick-Pt regions, revealing a distinct spin spiral profile under the different bottom-anisotropy conditions.

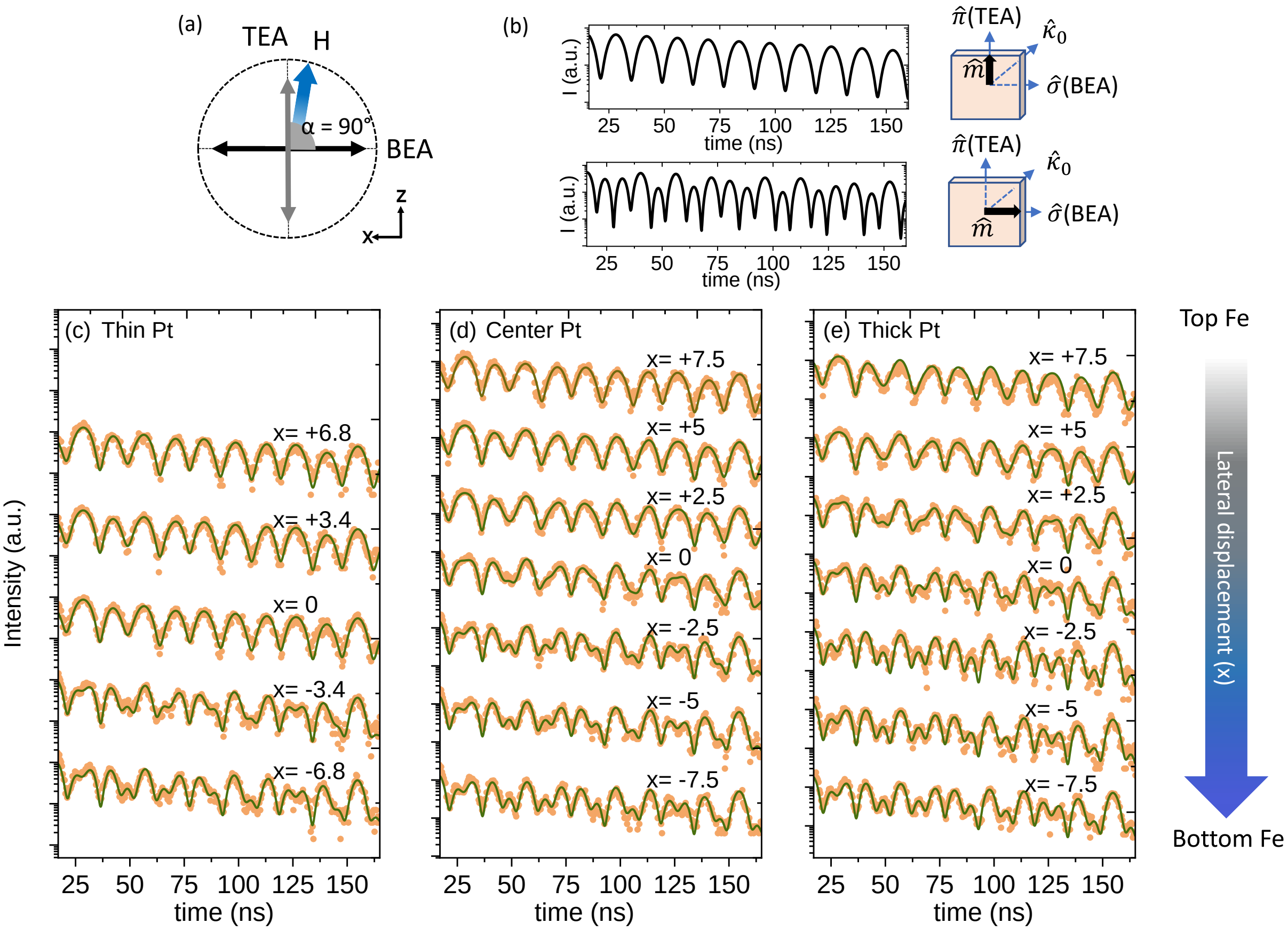

Fig. 6. Depth-dependent NFS quantum-beat patterns at remanence for three bottom-anisotropy conditions. (a) Schematic of the TEA–BEA configuration and the applied saturation field (H) oriented 5° relative to the TEA used to induce a vertical spin spiral at remanence. (b) Simulations of two NFS beat pattern assuming the total magnetization vector ($\vec{m}$) pointing in the TEA and BEA-direction, respectively. The beat patterns change when $\vec{m}$ rotates from TEA to BEA. (c-e) Experimental and fitted NFS time spectra measured at different x-positions, corresponding to different $^{57}$Fe probe-layer

depths from the top toward the bottom of the Fe film, for (c) the thin-Pt region (5.4 nm), (d) the center-Pt region (8 nm), and (e) the thick-Pt region (10.4 nm). Symbols denote the experimental data and solid lines the corresponding fits. The fitted spin-rotation angles are presented in Fig. 7.

For quantitative determination of the magnetization rotation angles through the Fe film thickness, the NFS time spectra were fitted using the model described in the Methods. The fitted hyperfine parameters and magnetization rotation angles are summarized in Supplementary Table 1. The extracted in-plane magnetization rotation angles as a function of depth are presented for all three Pt regions in Fig. 7(a), together with representations of the corresponding spin profiles in Fig. 7(b). The corresponding total spin-rotation angles are approximately 42°, 57° and 51° for the thick-, center-, and thin-Pt regions, respectively. Beyond these differences in total rotation, the fitted profiles show that the spin rotation is distributed differently through the film thickness across the three bottom-anisotropy conditions.

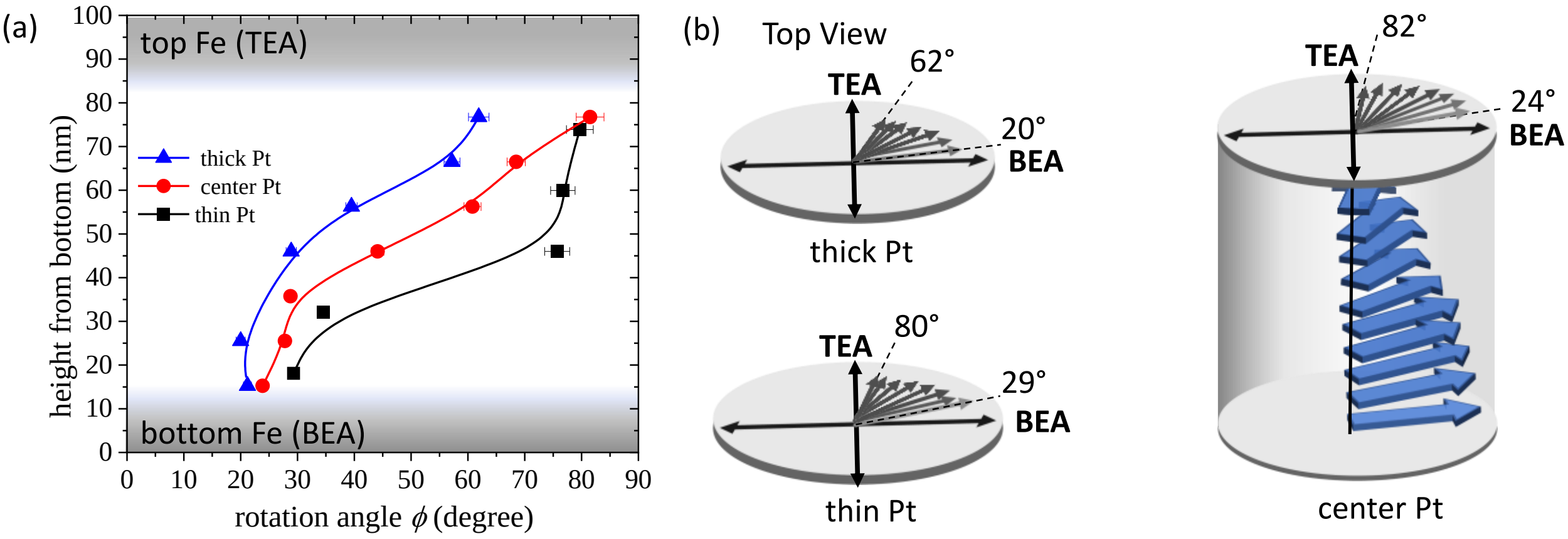

Fig. 7. Depth-dependent spin-spiral profiles obtained from NFS measurements depending on the bottom uniaxial anisotropy. (a) Depth dependence of the in-plane magnetization rotation angle for the thin-, center-, and thick-Pt regions, corresponding to different bottom-interface anisotropy strengths. The rotation angles were obtained from fits to the NFS time spectra measured at remanence. Symbols represent the fitted values, and solid lines are guides to the eye. The shaded area indicates the part of the sample that is not probed by the isotope layer. The plotted depth is assigned to the center of each interval; accordingly, the first and last data points correspond to the centers of the 10–20 nm and 70–80 nm intervals, respectively. (b) Top-view representations of the spin distributions through the film thickness for the thin- and thick-Pt regions, together with a three-dimensional representation of the depth-dependent magnetization profile for the center-Pt region.

## 3. Discussion

The depth-dependent spin profiles obtained from NFS show that a variation of the bottom-interface anisotropy modifies both the total spin rotation and its distribution through the film thickness. In particular, the thick-Pt region exhibits a smaller total rotation angle (~42°) than the thin-Pt region (~51°), whereas the intermediate center-Pt condition produces the largest rotation angle (~57°) and the most nearly uniform spin rotation through the film thickness. These differences can be understood in terms of the competition among exchange, anisotropy, and magnetostatic energy. The periodic surface corrugations generate spatially modulated magnetic surface charges and associated demagnetizing fields. [31–33]In the present double-sided OID film, both the top and bottom corrugated interfaces therefore contribute to the depth-dependent magnetostatic energy profile, while their crossed easy-axis orientations impose competing magnetic boundary conditions and anisotropies.

At the bottom interface we observe the influence of the OID-induced anisotropy tuning – with increasing anisotropy (from thin to thick Pt), the magnetization fixation gets stronger to the bottom easy axis and the rotation starts at a lower angle away from this axis. The rotation within the spiral is an interplay of the magnetostatic energy, exchange and shape-induced anisotropy contributions. The intrinsic magnetic exchange constant *A* of iron is expected to remain approximately unchanged on top of the varying Pt regions, whereas changing the Pt thickness modifies the wavy morphology and the corresponding anisotropy condition at the bottom interface, which is described by the effective anisotropy constant $K_{eff}$. This modulation is not only expected to change the BEA strength but also its extension into the film as the stray field conditions are altered with increasing correlated roughness. In total an anisotropy profile away from the bottom into the iron layer is expected influencing the total spin profile. The balance between exchange, which penalizes rapid spatial variations of the magnetization, and the anisotropy and magnetostatic contributions, which favor particular magnetization configurations, therefore changes across the Pt wedge. Within a simplified effective-anisotropy description, the competition between exchange and the combined anisotropy and magnetostatic contributions can be associated with a characteristic length scale proportional to $\sqrt{A/K_{eff}}$. A smaller anisotropy corresponds to a more gradual variation of the magnetization over a larger distance, whereas a larger anisotropy allows the magnetization to respond to the local anisotropy landscape over a shorter distance. [34,35] This holds for both interfaces and the spin spiral profile is therefore largely determined the anisotropies induced from both surface morphologies and stray field extension into the film.

The top OID iron film inducing the TEA is nominally the same over the sample. However, we observe a strong change of the rotation angle in the top part of the film with the bottom Pt thickness. Especially, for the thick OID Platinum layer on the bottom, we identify a significantly reduced coupling of the spins to the top easy axis. This effect could be due to correlated roughness. Even though, the 92 nm thick iron film of the stack, deposited in normal incidence, is expected to damp the wavy surface profile induced on the bottom by the OID Pt layer, it seems, that a remaining fraction with reduced amplitude is still transferred to the top side. As a consequence, the strength of anisotropy of the crossed top easy axis will be reduced leading to larger rotation away from the TEA axis. For the center and thin Pt thicknesses this disturbance of the TEA seems not to be present as their top rotations are similar.

The balance of the BEA and TEA and the influence of transferred roughness correlation during growth provides a strong handle to tune the spin spiral profile as seen in Fig. 7. Notably, the center-Pt region exhibits the most nearly uniform depth-dependent rotation, suggesting that the intermediate bottom-anisotropy condition provides a closer balance between the competing boundary constraints and the exchange interaction. Moving toward either weaker or stronger bottom-anisotropy conditions redistributes the spin rotation more unevenly through the film thickness.

The complementary sensitivities of PNR and NFS provide different views of the spin-spiral state. PNR probes the laterally in-plane magnetization profile through the full film thickness, whereas isotope-selective NFS provides sensitivity to the depth-resolved local hyperfine magnetic field and magnetization orientation at the probe layer, including sensitivity to out-of-plane magnetization components. PNR and NFS therefore complement each other. Together, the two techniques provide consistent evidence for the stabilization of a depth-dependent spin spiral at remanence.

Although the center-Pt region was designed with the same Pt thickness (8 nm) as the PNR spin-spiral sample and corresponds to a comparable calibrated bottom-anisotropy condition, the PNR and NFS measurements were performed on related but not magnetically identical specimens. Differences in capping layers, fabrication conditions, easy-axis separations, and magnetic field geometries can therefore contribute to differences between the obtained spin profiles. In particular, the saturation field was applied close to the TEA in the NFS, whereas in the PNR it was applied 35° relative to the TEA. The two profiles should therefore not be expected to coincide quantitatively, but instead provide complementary demonstrations of spin-spiral formation with different crossed magnetic anisotropies. Nevertheless, we get comparable spin spiral profiles in the PNR sample and the thick Pt sample of the NFS investigation showing that the spin spiral is sensitive to the exact sample and magnetic hysteresis.

Taken together, the results show that the crossed surface anisotropies not only stabilize a spin spiral at remanence but also provide control over its depth-dependent profile. This sensitivity to the anisotropies suggests that spiral configurations can be extended further by tuning both the strengths and relative orientations of the two anisotropies. OID deposition angle and layer thickness modify the surface morphology and associated anisotropy, and provide routes for extending the anisotropy control. For the experimentally realized spin spirals, micromagnetic simulations give a stored energy density of approximately 1.3 kJ $m^{-3}$ for a rotation angle of about 50° (see Supplementary Information). The same simulations predict that switching between wider spiral states (e.g. with opening angles between 20° and 160°), could increase the stored energy density to approximately 7 kJ $m^{-3}$.

The present results establish a single-film approach for stabilizing and tailoring vertical spin spirals in polycrystalline ferromagnetic Fe thin films at room temperature and at remanence using double-sided oblique-incidence deposition. By encoding crossed uniaxial anisotropies at the two opposing interfaces of a continuous ferromagnetic layer, both the orientation and relative strength of the magnetic boundary conditions can be used as design parameters for the resulting depth-dependent spin profile. The broader advance is therefore not merely the stabilization of a remanent spin spiral, but a route toward deliberately engineering the internal non-collinear magnetization profile of a conventional ferromagnetic film through its boundary conditions. This avoids the need for chemically distinct magnetic multilayers, complex interlayer-coupling architectures, cryogenic conditions, or Dzyaloshinskii–Moriya interaction.

## 4. Experimental section

**Sample Preparation:** Thin films and multilayers were prepared using a custom-made ultrahigh vacuum magnetron sputtering chamber with base pressure around $5 \times 10^{-8}$ mbar. The chamber is equipped with a rotatable sample stage and high-purity 1.5-inch targets of $^{nat}Fe$, $^{57}Fe$, Pt and Cr. For PNR measurements the sample structure is Si (001) substrate/OID $Pt_{80^\circ}$(8 nm)/$^{nat}Fe$ (92nm)/OID $^{nat}Fe_{80^\circ}$(8 nm)/Cr cap (8 nm) [Fig. 3(c)]. For NFS measurements, the stack of the layers as shown in Fig. 5 (b), is given as Si (001) substrate/OID $Pt_{80^\circ}$-wedge (5.4 nm-10.4 nm)/$^{nat}$ $Fe_{0^\circ}$ wedge (0-82 nm)/$^{57}Fe$ (10 nm)/ $^{nat}$ $Fe_{0^\circ}$wedge (82-0 nm)/OID $^{nat}Fe_{80^\circ}$ (8 nm)/Pt cap (8 nm), where the Pt cap was deposited on top of the sample to avoid the surface oxidation. A triangular shaped aperture was installed between the source and the substrate to produce a thickness gradient for $^{nat}Fe$ wedge layer (0-82 nm). The inclusion of a probe layer $^{57}Fe$ (10 nm) does not affect the magnetic state of the sample. We separately calibrated the thickness gradient of the Fe wedge by performing x-ray reflectometry (XRR) measurements at different regions on the sample.

**Magneto-optical Kerr effect and atomic force microscopy:** The magnetic anisotropy of the thin films was studied by MOKE measurements performed in longitudinal geometry using a He-Ne laser ($\lambda$ = 632 nm). The magnetic field was applied along the easy and hard axis in the plane of the film. AFM studies were carried out with a Brucker Dimension Icon AFM instrument to study the surface morphology of the OID induced waviness such as their periodicity, amplitude etc.

**Polarized neutron reflectometry:** PNR was performed at D17 beamline of the Institute Laue-Langevin[36] in Grenoble, France.[37] The measurements were performed in horizontal scattering plane geometry with vertical sample alignment in polarized time-of-flight mode (ToF) with a wavelength band of 0.4 - 2 nm. The neutrons were polarized parallel to the x direction as shown schematically in the scattering geometry in Fig. 4(a). The intensity of the reflected neutrons (R) and their spin dependence were recorded as a function of the scattering vector $q_z = (4\pi/\lambda)\sin\theta$ (where $\theta$ is incident angle and $\lambda$ is the wavelength of the incident neutrons).[38] $R^{++}$ and $R^{--}$denote the non-spin-flip (NSF) scattering channels that measure the reflected intensities where the polarized neutrons do not change their spin state during the scattering process. The NSF intensity probes the nuclear scattering length density profile and the component of the magnetization in the sample that is aligned parallel or antiparallel to the applied guide field that was oriented along the x-axis. On the other hand, in the spin-flip (SF) scattering channels denoted as $R^{+-}$ and $R^{-+}$, the reflected intensities of the neutrons are detected with changing spin states after the scattering process. Here, the SF scattering arises from magnetic interactions only and measures finite projections of the magnetization perpendicular to the neutron polarization. Fitting of the PNR data was carried out using GenX software.[39] The remanent PNR data show unequal $R^{+-}$ and $R^{-+}$ spin-flip intensities. Therefore, only the SF ($R^{+-}$) and both NSF ($R^{++}$ and $R^{--}$) channels of the reflectivity were fitted. The high-field PNR spectra were fitted first, with the Fe layer divided into 10 uniformly magnetized sublayers aligned with the applied field. The structural parameters and saturation magnetization obtained from the high-field fit were then used as starting values for fitting the PNR data measured at remanence under a 0.5 mT guiding field. In the remanent fit, the saturation magnetization and the magnetization orientation of the individual Fe sublayers were fitted to obtain the spin-spiral profile.

**Nuclear forward scattering:** Nuclear forward scattering of synchrotron radiation is the time-analogue of Mössbauer spectroscopy, which probes the hyperfine interactions of nuclear transitions in $^{57}$Fe, via nuclear resonant scattering of 14.4 keV photons. The decay of the excited nucleus shows quantum beats in the temporal evolution of the nuclear resonant scattered intensity.[15,19]NFS is able to investigate the magnetic moment orientation (in-plane and out-of-plane) with an accuracy of around 1° in magnetic materials.[40] Careful analysis of these temporal beat patterns gives information about the strength and orientation of the magnetization ($\vec{m}$) relative to the photon wavevector ($k_0$) and its polarization base ($\vec{\sigma}$,$\vec{\pi}$). Nuclear resonance scattering in reflection geometry is a standard approach for thin films and is widely used. However, for 100 nm thick films, the sensitivity to the bottom layers would be reduced due to the limited x-ray penetration at grazing incidence. To overcome this, we implemented a $^{57}$Fe probe layer approach and performed the measurements in transmission geometry shown in Fig. 5 (c) at the dynamics beamline P01 at PETRA III, Hamburg, Germany.[41] The count rate of the delayed photons was 2-3 Hz. To probe the three Pt regions (thin, center, and thick), the sample was vertically displaced along the z-axis, while lateral displacements along the x-axis (perpendicular to $\overrightarrow{k_0}$) in 2.5 mm steps (x = –7.5 mm to +7.5 mm) enabled depth-resolved screening of the $^{57}$Fe probe layer within the 100 nm Fe film sample across the 15 mm-long sample.

The fitting of the recorded time spectra was carried out using the NEXUS software[42] with the following procedure. We started with the fitting of in-field time spectra measured at 230 mT at the central position, where the magnetic field was applied along the BEA. For fitting of the in-field time spectra, the 100 nm Fe layer was divided into total ten sub-layers of which nine corresponded to $^{nat}$Fe and one to $^{57}$Fe, with all the layers oriented along $k_0$. From the fit we obtained the hyperfine field parameters of all the layers. After determining the hyperfine-field parameters, we proceeded to fit the time spectra in the remanent state. We used a common fit model for all spectra in which we introduced a spin spiral (systematic variation of the spin angles in all the layers from the top towards the bottom depth) in the background (i.e., for nine $^{nat}$Fe layers) except for one $^{57}$Fe layer. The initial guess of the spin angles for the layers was based on the PNR results. To precisely determine the spin rotation angle of the background spin spiral ($^{nat}$Fe), we first applied this fit model to the extreme x-positions (i.e., at x= +7.5 and x = -7.5) for the center-Pt region, where the spins of $^{57}$Fe could be assumed to be aligned close to the top and bottom easy axes, respectively. After determining the hyperfine parameters (especially the spin-rotation angles) of the background spiral, we fitted the in-plane and out-of-plane orientations of the spin moments of the $^{57}$Fe layer at different depths. The resulting fit parameters then provided guidance for fitting the time spectra in thin-Pt and thick-Pt regions using the same approach. Note that the natural abundance of $^{57}$Fe is only 2 % and the signal from the non-enriched film corresponds to 15 % of the signal.

## Acknowledgments

We acknowledge DESY (Hamburg, Germany; Helmholtz Association HGF) for supporting our experiments by providing access to their facilities. We also acknowledge the Institut Laue–Langevin (ILL) and the Australian Nuclear Science and Technology Organization (ANSTO) for enabling measurements at their PNR facilities. Beamtimes were allocated under proposals Nos. I-20230542 (DESY), 5-54-401 (ILL), and 17204 (ANSTO). We thank Dr. Michele Buzzi, Condensed Matter Department, Max Planck Institute for the Structure and Dynamics of Matter, Hamburg, Germany, for the SQUID measurements. This work was supported by the Cluster of Excellence "CUI: Advanced Imaging of Matter" of the Deutsche Forschungsgemeinschaft (DFG), EXC 2056, project ID 390715994.

## Bibliography

1. Yang, S.-H., Naaman, R., Paltiel, Y. & Parkin, S. S. P. Chiral spintronics. *Nature Reviews Physics* **3**, 328–343 (2021).

2. Jeon, Y. S. *et al.* Spin-selective transport through chiral ferromagnetic nanohelices. *Science (1979).* **389**, 1031–1036 (2025).

3. Kim, Y.-H. *et al.* Chiral-induced spin selectivity enables a room-temperature spin light-emitting diode. *Science (1979).* **371**, 1129–1133 (2021).

4. Fernández-Pacheco, A. *et al.* Three-dimensional nanomagnetism. *Nat. Commun.* **8**, 15756 (2017).

5. Sanz-Hernández, D. *et al.* Artificial double-helix for geometrical control of magnetic chirality. *ACS Nano* **14**, 8084–8092 (2020).

6. Menzel, M. *et al.* Information Transfer by Vector Spin Chirality in Finite Magnetic Chains. *Phys. Rev. Lett.* **108**, 197204 (2012).

7. Phatak, C. *et al.* Visualization of the magnetic structure of sculpted three-dimensional cobalt nanospirals. *Nano Lett.* **14**, 759–764 (2014).

8. Khajetoorians, A. A., Wiebe, J., Chilian, B. & Wiesendanger, R. Realizing all-spin–based logic operations atom by atom. *Science (1979).* **332**, 1062–1064 (2011).

9. Vedmedenko, E. Y. & Altwein, D. Topologically protected magnetic helix for all-spin-based applications. *Phys. Rev. Lett.* **112**, 17206 (2014).

10. Parkin, S. S. P., More, N. & Roche, K. P. Oscillations in exchange coupling and magnetoresistance in metallic superlattice structures: Co/Ru, Co/Cr, and Fe/Cr. *Phys. Rev. Lett.* **64**, 2304 (1990).

11. Stiles, M. D. Interlayer exchange coupling. *J. Magn. Magn. Mater.* **200**, 322–337 (1999).

12. Parkin, S. S. P. Systematic variation of the strength and oscillation period of indirect magnetic exchange coupling through the 3d, 4d, and 5d transition metals. *Phys. Rev. Lett.* **67**, 3598 (1991).

13. Paul, A. Stiffness in vortex—like structures due to chirality-domains within a coupled helical rare-earth superlattice. *Sci. Rep.* **6**, 19315 (2016).

14. Röhlsberger, R. *et al.* Imaging the magnetic spin structure of exchange-coupled thin films. *Phys. Rev. Lett.* **89**, 237201 (2002).

15. Fust, S. *et al.* Realizing topological stability of magnetic helices in exchange-coupled multilayers for all-spin-based system. *Sci. Rep.* **6**, 1–14 (2016).

16. Ye, J., Baldauf, T., Mattauch, S., Paul, N. & Paul, A. Topologically stable helices in exchange coupled rare-earth/rare-earth multilayer with superspin-glass like ordering. *Commun. Phys.* **2**, (2019).

17. Huang, J. *et al.* Non-collinear magnetic configuration mediated exchange coupling at the interface of antiferromagnet and rare-earth nanolayers. *Sci. Rep.* **12**, (2022).

18. Dzemiantsova, L. V, Meier, G. & Röhlsberger, R. Stabilization of magnetic helix in exchange-coupled thin films. *Sci. Rep.* **5**, 16153 (2015).

19. Schlage, K. & Röhlsberger, R. Nuclear resonant scattering of synchrotron radiation: Applications in magnetism of layered structures. *J. Electron Spectros. Relat. Phenomena* **189**, 187–195 (2013).

20. Liedtke, S., Grüner, C., Lotnyk, A. & Rauschenbach, B. Glancing angle deposition of sculptured thin metal films at room temperature. *Nanotechnology* **28**, (2017).

21. Singh, S. *et al. Tailoring Magnetic Properties of Zigzag Structured Thin Films via Interface Engineering and Columnar Nano-Structuring*.

22. Bera, A. K. *et al.* Morphology induced large magnetic anisotropy in obliquely grown nanostructured thin film on nanopatterned substrate. *Appl. Surf. Sci.* **581**, (2022).

23. Bubendorff, J. L. *et al.* Origin of the magnetic anisotropy in ferromagnetic layers deposited at oblique incidence. *Europhys. Lett.* **75**, 119–125 (2006).

24. Willing, S. *et al.* Novel Tunnel Magnetoresistive Sensor Functionalities via Oblique-Incidence Deposition. *ACS Appl. Mater. Interfaces* **13**, 32343–32351 (2021).

25. Kueny, E. *et al.* Spin-structured multilayer THz emitters by oblique incidence deposition. *J. Appl. Phys.* **133**, (2023).

26. Barranco, A., Borras, A., Gonzalez-Elipe, A. R. & Palmero, A. Perspectives on oblique angle deposition of thin films: From fundamentals to devices. *Prog. Mater. Sci.* **76**, 59–153 (2016).

27. Suzuki, M. Practical applications of thin films nanostructured by shadowing growth. *J. Nanophotonics* **7**, 073598 (2013).

28. Schlage, K. *et al.* Spin-Structured Multilayers: A New Class of Materials for Precision Spintronics. *Adv. Funct. Mater.* **26**, 7423–7430 (2016).

29. Willing, S., Oepen, H. P. & Roehlsberger, R. *Oblique-Incidence Deposition of Ferromagnetic Thin Films and Their Application in Magnetoresistive Sensors*. (Uni Hamburg/Experimentalphysik, 2020).

30. Schlage, K. *et al.* The formation and magnetism of iron nanostructures on ordered polymer templates. *New J. Phys.* **14**, (2012).

31. Schlömann, E. Demagnetizing fields in thin magnetic films due to surface roughness. *J. Appl. Phys.* **41**, 1617–1622 (1970).

32. Liedke, M. O. *et al.* Crossover in the surface anisotropy contributions of ferromagnetic films on rippled Si surfaces. *Phys. Rev. B Condens. Matter Mater. Phys.* **87**, (2013).

33. Arranz, M. A., Colino, J. M. & Palomares, F. J. On the limits of uniaxial magnetic anisotropy tuning by a ripple surface pattern. *J. Appl. Phys.* **115**, (2014).

34. Thiaville, A., Rohart, S., Jué, É., Cros, V. & Fert, A. Dynamics of Dzyaloshinskii domain walls in ultrathin magnetic films. *EPL* **100**, (2012).

35. Hellman, F. *et al.* Interface-induced phenomena in magnetism. *Rev. Mod. Phys.* **89**, (2017).

36. Panchwanee, A., Bocklage, L., Lott, D., Saerbeck, T. & Schlage, K. *Probing Depth Dependent Spiral Spin Structures in Obliquely Deposited FeCo Thin Films*. https://doi.org/10.5291/ILL-DATA.5-54-401 (2023).

37. Saerbeck, T. *et al.* Recent upgrades of the neutron reflectometer D17 at ILL. *J. Appl. Crystallogr.* **51**, 249–256 (2018).

38. Causer, G. L., Guasco, L., Paull, O. & Cortie, D. Topical Review of Quantum Materials and Heterostructures Studied by Polarized Neutron Reflectometry. *Physica Status Solidi - Rapid Research Letters* vol. 17 Preprint at https://doi.org/10.1002/pssr.202200421 (2023).

39. Glavic, A. & Björck, M. GenX 3: The latest generation of an established tool. *J. Appl. Crystallogr.* **55**, 1063–1071 (2022).

40. Röhlsberger, R. *Nuclear Condensed Matter Physics with Synchrotron Radiation: Basic Principles, Methodology and Applications*. vol. 208 (Springer Science & Business Media, 2004).

41. Wille, H. C., Franz, H., Röhlsberger, R., Caliebe, W. A. & Dill, F. U. Nuclear resonant scattering at PETRA III: Brillant opportunities for nano–and extreme condition science. in *Journal of Physics: Conference Series* vol. 217 12008 (2010).

42. Bocklage, L. Nexus - Nuclear Elastic X-ray scattering Universal Software. Preprint at https://doi.org/10.5281/zenodo.7716208 (2023).

**Supplement material for**

# Customized spin spirals in ferromagnetic thin films

Anjali Panchwanee,[1, a)] Kai Schlage,[1, b)] Dieter Lott,[2] Sven Velten,[1] Thomas Saerbeck,[3] David L. Cortie,[4] Sakshath Sadashivaiah,[5] Ilya Sergeev,[1] Guido Meier,[6, 7] Lars Bocklage,[1, 7] and Ralf Rohlsberger[1, 5, 7, 8, 9]

1) Deutsches Elektronen-Synchrotron DESY, Notkestraße 85, 22607 Hamburg, Germany

2) Institute for Materials Research, Helmholtz-Zentrum Geesthacht, Max-Planck-Straße 1, 21502 Geesthacht, Germany

3) Institut Laue-Langevin, 71 avenue des Martyrs, CS 20156, 38042 Grenoble cedex 9, France

4) Australian Nuclear Science and Technology Organisation, New Illawarra Road, Lucas Heights, NSW 2234, Australia

5) Helmholtz-Institut Jena, Fröbelstieg 3, 07743 Jena, Germany

6) Max Planck Institute for the Structure and Dynamics of Matter, Luruper Chaussee 149, 22761 Hamburg, Germany

7) The Hamburg Centre for Ultrafast Imaging, Luruper Chaussee 149, 22761 Hamburg, Germany

8) Institut für Optik und Quantenelektronik, Friedrich-Schiller-Universität Jena, Max-Wien-Platz 1, 07743 Jena, Germany

9) GSI Helmholtzzentrum für Schwerionenforschung GmbH, Planckstraße 1, 64291 Darmstadt, Germany

## Atomic force microscopy

To demonstrate that the thickness of the OID layer controls the amplitude of the surface waviness, the surface topography was investigated using AFM. The AFM images of the OID Pt wedge thin film at the thick and thin Pt regions are presented in Fig. S1 (a, b). We observe several long, connected Pt columns oriented perpendicular to the Pt deposition direction (indicated by the arrow in the AFM image) in the thick Pt region while in the thinner regions these columns are less pronounced/not detectable? (d). This indicates a reduction in the strength of anisotropic surface morphology in the thin-Pt region relative to the thick-Pt region.

The line profiles of the AFM images, taken parallel to the ripple vector, are shown in Fig. S1 (a, b). The average ripple wavelength (λ) for the thick and thin Pt regions is 24 nm and 20 nm, respectively. The RMS roughness (average ripple amplitude) of the rippled patterns, estimated from 1 μm x 1 μm AFM images, is 1.8 nm and 1.3 nm for the thick and thin-Pt regions, respectively. As a result, this leads to a reduced strength of shape-induced uniaxial anisotropy in the Fe layer grown on top of the thin-Pt region compared to the thick-Pt region, as later confirmed by MOKE measurements shown in Fig. S2. Similarly, the surface topography of the 8 nm OID iron layer deposited on the Si substrate and the corresponding line profile are shown in Fig. S1(c). The 3D surface topography of the same film is shown in Fig. S1(d). The obtained ripple wavelength and RMS roughness (ripple amplitude) are 19.6 nm and 0.9 nm, respectively, indicating that the associated strength of the shape-induced uniaxial magnetic anisotropy is smaller than that of Fe, grown on the OID thick-Pt.

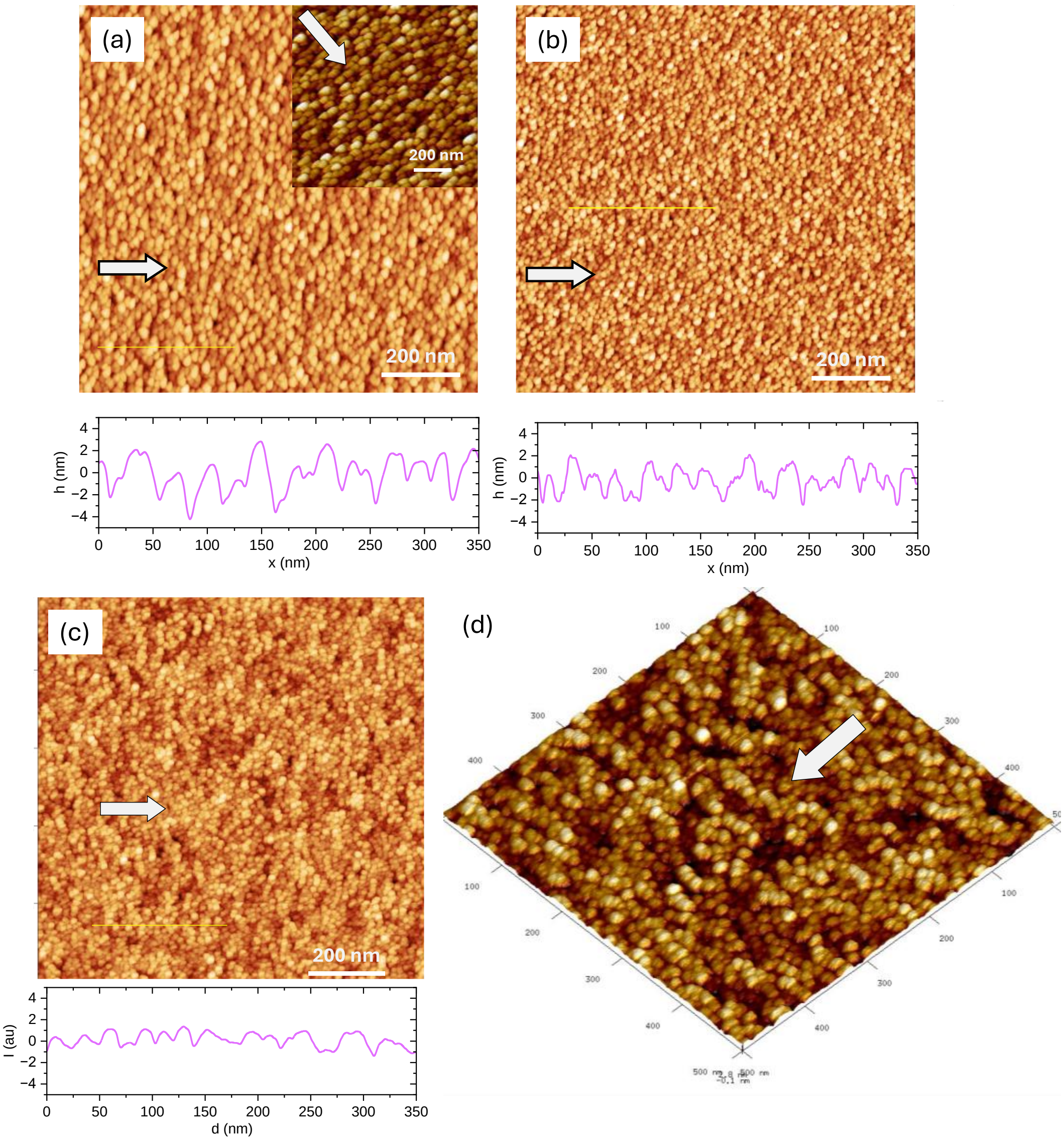

Fig. S1. Atomic force microscopy (AFM) images of the OID Pt wedge thin film at (a) thick and (b) thin Pt regions along with the line profiles shown by yellow line on the image. (c) AFM image and line profile of 8 nm OID iron layer and (d) its 3 D topographic view. Inset of (a) is 3D topographic view of thick Pt region.

**Transferred magnetic anisotropy strengths:**

By varying the amplitude of the wavy surface profile of the bottom OID Pt layer (θ = 80˚) through changes in its thickness, we can effectively tune the strength of the uniaxial magnetic anisotropy in the top Fe layer (t = 5 nm). Figs. S2(b) and S2(c) present corresponding hysteresis loops of the iron layer for OID Pt film thicknesses ranging from 1.5 nm to 10.4 nm, demonstrating a systematic increase in the anisotropy strength with increasing Pt thickness ($t_{Pt}$). We determined the uniaxial anisotropy strength as a function of Pt thickness. The uniaxial anisotropy energy density[1] is given by

$$K_u = \frac{1}{2}\mu_0 M_s H_k \qquad \text{Eq. (1)}$$

where, $\mu_0$ is the vacuum permeability, $M_s$ is the saturation magnetization (1.7x10$^6$ A/m for a polycrystalline Fe thin film[1]), and $H_k$ is the anisotropy field. $K_u$ is then determined by fits of the MOKE data in Fig. S2 (c) with an arctan function.[2] Its dependence on the Pt thickness is shown in Fig. S2(d). The linear increase with thickness demonstrates that the amplitude of the waviness of the underlying OID Pt layer directly governs the magnetic properties of the Fe layer, allowing controlled tuning of its uniaxial anisotropy strength.

In the magnetizations of Fe thin film grown on OID Pt, we observed a reduction of up to 10% in the remanent magnetization for $t_{Pt} \geq 8$ nm, indicating the emergence of an out-of-plane magnetic contribution. We noted that the out of plane magnetization contribution is very difficult to estimate from MOKE measurements in the spiral sample due to the shape of MOKE hysteresis loop.

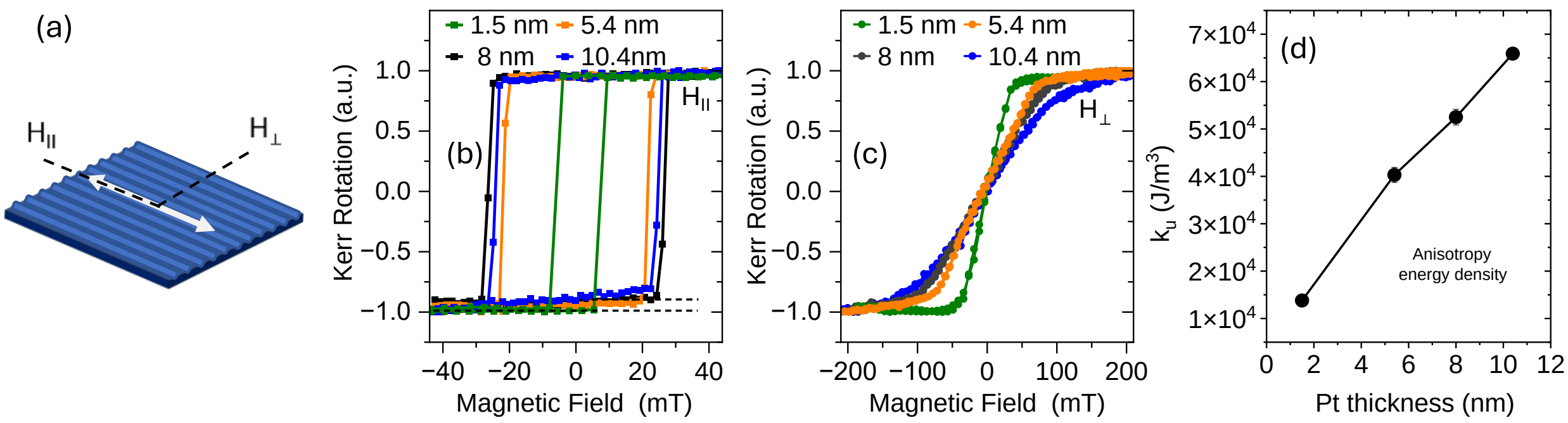

Fig. S2. Tuning of the anisotropy strength of the bottom easy axis in the spiral sample. (a) schematic of a Fe (5 nm) film on top of Pt (t = 1.5, 5.4, 8, 10.4 nm) OID at θ = 80°. The induced easy axis is shown by the white bidirectional arrow. (b-c) Normalized in-plane MOKE hysteresis loops under an in-plane applied magnetic field parallel (∥) and perpendicular (⊥) to the induced in-plane easy axis, as a function of different Pt thicknesses ($t_{Pt}$) of the bottom OID-Pt layer, respectively. (d) Variation of the anisotropy energy density ($k_u$) as a function of different bottom $t_{Pt}$. The solid line is a guide to the eye and $k_u$ is estimated using Eq. (1).

**SQUID measurements:**

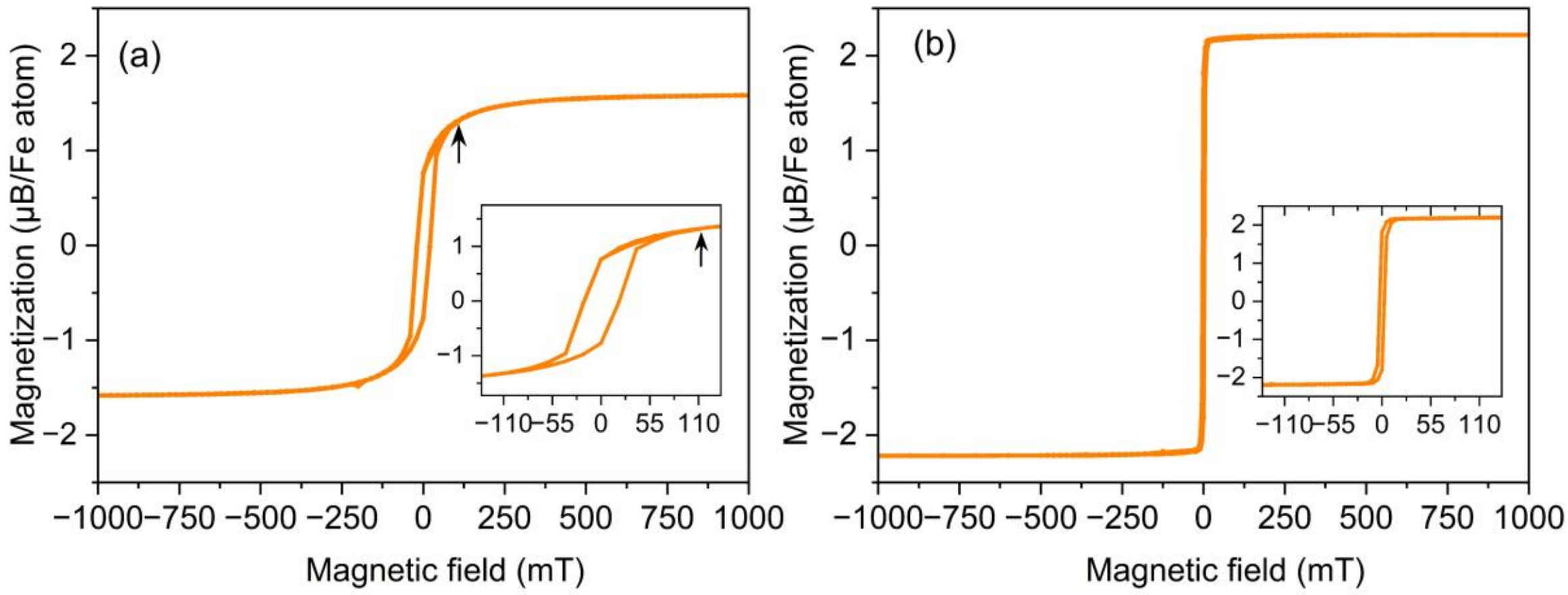

Fig.S3 Room-temperature SQUID magnetization hysteresis (M–H) loops measured for (a) PNR sample and (b) NFS sample at the central Pt-thickness region.

**Time spectra fitting results from NFS measurements:**

**Table 1:** Hyperfine parameters extracted from the fitting of the time spectra for the three-thickness of OID Pt wedge film at thin, center and thick corresponding to different anisotropy strength of BEA. The in-plane magnetization angle, $\phi$ (deg), varies with x-displacement position, while the out-of-plane angle was fixed at (θ = 1.8°). The hyperfine field, BHF, was approximately 32.8–32.9 T, with a common hyperfine-field distribution of 0.83 T.

| **(a) Thin-Pt** | |
|---|---|
| x-displacement (mm) | $\phi$ (deg) (in-plane rotation angle) |
| +6.8 | 79.7±2.4 |
| +3.4 | 76.74±2 |
| 0 | 75.76±2.2 |
| -3.4 | 34.46±0.98 |

| -6.8 | 29.27±0.8 |
|---|---|
| **(b) Center-Pt** | |
| +7.5 | 81.54±2.4 |
| +5 | 68.54±1.6 |
| +2.5 | 60.85±1.5 |
| 0 | 44.1±1 |
| -2.5 | 28.76±0.8 |
| -5 | 27.8±0.8 |
| -7.5 | 23.87±0.9 |
| **(c) Thick-Pt** | |
| +7.5 | 62±1.8 |
| +5 | 57.2±1.4 |
| +2.5 | 39.55±1 |
| 0 | 28.95±0.9 |
| -2.5 | 15.65±0.86 |
| -5 | 20±0.8 |
| -7.5 | 21.23±0.7 |

**Estimation of energy storage density in spiral state via micromagnetic simulations:**

Micromagnetic simulations were carried out using the MicroMagnum code[3] to obtain the stabilized equilibrium state via energy minimization and to estimate the stored energy density in the spin spiral. This approach calculates the Landau-Lifshitz-Gilbert (LLG) equation[4] and the effective field ($H_{eff}$) in the LLG equation which is the sum of magnetic exchange energy, anisotropy energy, demagnetization energy, and external magnetic field energy. Based on these simulations, we observed that the estimated internal effective energy in the spiral state is different than the uniaxial / collinear ferromagnetic state. We estimated the effective magnetic energy density E stored in the spin spirals for different relative angle with the top and the bottom easy axes (α). The estimated stored energy density in our stabilized spiral for the opening angle of 50˚ with respect to the 0˚ spiral is E ~ 1.3 kJ/m$^3$. The stored energy in the spiral can further be increased by increasing the angle (α) between the crossed TEA and BEA as discussed previously in the sample fabrication. For example, for α = 160˚, in zero field, two equilibrium spiral states corresponding to opening angle of 160˚ and 20˚ respectively can be stabilized, which depends on the orientation of the starting magnetization direction in the sample. The energy stored in such a spiral is 7 kJ/m$^3$.

In the present calculation, we used 100 nm thick Fe which is divided into 2×1×100 cells along *x*×*y*×*z* direction, and using period boundary conditions in *xy* direction. Size of one cell is 1×1×1 nm$^3$. For Fe layer, we used $M_s$ = 1.7 x 10$^6$ A/m, exchange constant (A) = 1.0 x 10$^{-11}$ J/m,[5] Gilbert damping factor = 0.01. However, in the present case, surface anisotropy contributions (from top and bottom parts) have to be considered in order to obtain the equilibrium spin configuration. Therefore, we have divided Fe thickness into three magnetic layers such as bottom, intermediate and top indicated by $Fe_b$, $Fe_{int}$, and $Fe_t$, respectively. We used the following parameters for the micromagnetic simulations:

| α = 90˚/ α = 0 ˚ | | | α = 160˚ / α = 20˚ |
|---|---|---|---|
| layer | Thickness (nm) | $K_u$ (J/m$^3$) | $K_u$ (J/m$^3$) |
| $Fe_t$ | 3.5 | 6.5 x 10$^4$ | 5.0 x 10$^5$ |
| $Fe_{int}$ | 97 | 0 | 0 |
| $Fe_b$ | 3.5 | 8.0 x 10$^4$ | 4.0 x 10$^6$ |

**Table 2:** The parameters used to estimate the energy density stored E in the spiral using micromagnetic simulations for relative angle between the TEA and BEA (α). In all cases external magnetic field is set to zero. For α = 90˚/ α = 0˚, the orientation of the initial magnetic state is set to (1,1,0). For α = 160˚/ α = 20 ˚, initial magnetic state orientation (1,0,0) and (0,1,0) respectively.

**Bibliography**

1. Cullity, B. D. & Graham, C. D. *Introduction to Magnetic Materials*. (John Wiley & Sons, 2011).
2. Panchal, G., Choudhary, R. J., Kumar, M. & Phase, D. M. Interfacial spin glass mediated spontaneous exchange bias effect in self-assembled La0. 7Sr0. 3MnO3: NiO nanocomposite thin films. *J. Alloys Compd.* **796**, 196–202 (2019).
3. Kim, H. & You, C.-Y. Embedded Object-Oriented Micromagnetic Frame (OOMMF) for More Flexible Micromagnetic Simulations. *Journal of Magnetics* **21**, 491–495 (2016).
4. Lifshitz, E. M. & Pitaevskii, L. P. *Statistical Physics: Theory of the Condensed State*. vol. 9 (Elsevier, 2013).
5. Röhlsberger, R. *et al.* Imaging the magnetic spin structure of exchange-coupled thin films. *Phys. Rev. Lett.* **89**, 237201 (2002).